\documentclass[preprint,12pt]{elsarticle}

\usepackage{amssymb}
\usepackage{float}
\usepackage{color, colortbl}
\usepackage{lipsum} 
\usepackage{url}
\begin{document}

\begin{frontmatter}


\title{The JARVIS Infrastructure is All You Need for Materials Design}



\author[aff1,aff2,aff3]{Kamal Choudhary}
 \address[aff1]{%
 Material Measurement Laboratory, National Institute of Standards and Technology,
Gaithersburg, MD 20899, USA 
}
\address[aff2]{Department of Materials Science and Engineering, Whiting School of Engineering, The Johns Hopkins University, Baltimore, MD 21218, USA}
\address[aff3]{Department of Electrical and Computer Engineering, Whiting School of Engineering, The Johns Hopkins University, Baltimore, MD 21218, USA}

\begin{abstract}

The Joint Automated Repository for Various Integrated Simulations (JARVIS) is a unified platform for multiscale, multimodal, forward, and inverse materials design. It integrates diverse theoretical and experimental approaches, including density functional theory, quantum Monte Carlo, tight-binding, classical force fields, machine learning, microscopy, diffraction, and cryogenics, across a wide range of materials. Emphasizing open access and reproducibility, JARVIS provides datasets, tools, benchmarks, and web applications that are widely adopted by the materials community. By bridging computation and experiment, JARVIS accelerates both fundamental research and real-world materials innovation.

\end{abstract}




\end{frontmatter}

\tableofcontents
\newpage
\section{Introduction}
Materials design has long been a cornerstone of technological progress, underpinning advancements in energy, computing, healthcare, and manufacturing \cite{callister2020fundamentals,olson1997computational,rajan2015materials}. Traditionally, materials discovery followed an empirical, trial-and-error approach, often requiring decades to identify and optimize materials for specific applications. The advent of computational materials science, particularly with the rise of density functional theory (DFT), tight-binding (TB), molecular dynamics (MD)/force-field (FF) and machine learning (ML) has significantly accelerated the pace of discovery \cite{jain2013commentary,curtarolo2013high,choudhary2020joint}. However, the growing complexity of materials problems, the vast chemical and structural search space, and the increasing demand for high-performance materials necessitate more sophisticated, integrated frameworks that can streamline the entire process from fundamental property prediction to real-world implementation.

An integrated materials design infrastructure bridges the gap between theoretical models, computational simulations, and experimental validation, ensuring that materials research is reproducible, scalable, and accessible. Moreover, multiscale modeling \cite{tadmor2011modeling} capabilities available in these infrastructures enable realistic materials design. Such frameworks incorporate multimodal datasets, automated workflows, and ML models, enabling both forward design (predicting properties from structures) \cite{schmidt2019recent,choudhary2022recent,vasudevan2019materials,schleder2019dft} and inverse design (identifying structures with desired properties) \cite{saal2013materials,sanchez2018inverse,agrawal2016perspective,zunger2018inverse}. Forward design remains challenging due to the need for high-throughput calculations across multiple material classes, while inverse design requires advanced optimization techniques, generative models, and robust experimental validation.

One of the critical challenges in forward materials design is the computational cost associated with methods such as first-principles simulations. While DFT and beyond methods are widely used for predicting electronic and structural properties, they are limited by system size and accuracy trade-offs, particularly for strongly correlated materials and high-temperature phases \cite{sholl2022density}. In classical approaches, empirical force fields enable large-scale simulations but often lack transferability across different materials and conditions \cite{plimpton1995fast}. More recently, machine learning property prediction models and machine learning force fields (MLFFs) have emerged as a promising alternative, offering near-DFT accuracy at significantly lower computational cost, yet challenges remain in ensuring generalizability and robustness \cite{behler2007generalized,zhang2018deep}. For experimental methods, availability of standardized benchmarks, integration of analytical and computational methods at various lengths and time scales and framework to enhance reproducibility is still a critical challenge \cite{baker20161}. 

Inverse design, on the other hand, relies heavily on data-driven methodologies, requiring extensive databases of materials properties and efficient search algorithms \cite{zunger2018inverse,lee2023machine}. Traditional approaches, such as genetic algorithms and evolutionary strategies, have been complemented by deep learning models, which can generate candidate materials based on target functionalities \cite{gomez2018automatic}. However, a major bottleneck is the integration of experimental validation with computational predictions, as synthetic feasibility and stability assessments are often overlooked in algorithmic designs \cite{zhong2022explainable}. Additionally, the lack of standardized benchmarking protocols makes it difficult to compare different methods and validate new materials against experimental findings \cite{zeni2023mattergen}.

Given these challenges, an effective materials design infrastructure must not only provide computational tools but also enable reproducibility, findability, accessibility, interoperability, and reusability (FAIR) \cite{wilkinson2016fair} data practices, and seamless integration of theoretical, computational, and experimental methods. The development of platforms such as Joint Automated Repository for Various Integrated Simulations (JARVIS), which integrates quantum and classical calculations, machine learning models, and experimental datasets, addresses these gaps by offering a unified ecosystem for forward and inverse materials design. By standardizing workflows, automating simulations, and facilitating community-driven data sharing, such infrastructures pave the way for accelerated discoveries and technological breakthroughs in materials science.

The JARVIS infrastructure \cite{choudhary2020joint,wines2023recent} distinguishes itself from other materials design platforms by offering a multimodal, multiscale, and reproducible ecosystem that integrates first-principles calculations, machine learning models, experimental datasets, and interactive web-based tools into a unified framework.

\textcolor{black}{Prominent materials informatics resources such as the Automatic FLOW for Materials Discovery (AFLOW) \cite{curtarolo2012aflow}, Materials Project \cite{jain2013commentary}, Open Quantum Materials Database (OQMD) \cite{saal2013materials}, Novel Materials Discovery (NOMAD) \cite{draxl2018nomad}, Materials Cloud \cite{pizzi2016aiida}, and the Open Catalyst Project (OC20) \cite{chanussot2021open} have significantly advanced DFT-based materials screening. However, they typically focus on a single modeling scale and do not broadly incorporate methods such as tight-binding, classical force-fields, dynamical mean field theory (DMFT), quantum Monte Carlo (QMC), microscopy, cryogenic measurements, and quantum computing techniques. In contrast, JARVIS integrates these diverse modeling strategies to support a true multiscale design workflow. For example, JARVIS-QETB supports tight-binding parameterization; JARVIS-FF and ALIGNN-FF enable large-scale simulations with classical and machine-learned force fields; and JARVIS-QC explores quantum computation algorithms for materials modeling. Furthermore, JARVIS encompasses a wide range of material classes and applications, including superconductors, solar cells, thermoelectrics, piezoelectrics, dielectrics, low-dimensional materials, catalytic systems, topological insulators, and datasets for experimental validation.}

A key differentiator is reproducibility-JARVIS provides open-access, FAIR-compliant datasets and workflows distributed via web applications, notebooks, and the JARVIS-Leaderboard \cite{choudhary2024jarvis}, which facilitates rigorous benchmarking across multiple domains. This ensures that results can be independently verified and extended by the community. Finally, JARVIS is designed for scalability, offering not only high-throughput DFT and force-field data, but also AI-driven models (e.g., ALIGNN, AtomGPT) that can accelerate materials prediction at significantly reduced computational cost compared to conventional approaches \cite{zuo2020performance}.

Unlike various other platforms, which often focus on a single domain such as semiconductors, metals, or alloys, JARVIS encompasses a broad range of materials, including superconductors, carbon capture frameworks, and low-dimensional heterostructures. It supports both forward and inverse design approaches and integrates a diverse set of theoretical and experimental techniques-spanning quantum Monte Carlo, classical force fields, microscopy, and diffraction data to ensure that predictions are not only computationally robust but also experimentally validated. This multimodal integration bridges the gap between in silico discoveries and real-world applications.

While the prediction and identification of new compounds are central to JARVIS, the infrastructure also facilitates the full materials design process. It enables multiscale property prediction across quantum, classical, and experimental domains; supports targeted materials optimization using models such as ALIGNN and AtomGPT, often guided by leaderboard-based performance rankings; and promotes reproducibility and transparency through rigorous benchmarking protocols. Workflow automation, enabled by the JARVIS-tools Python package, ensures consistent and reproducible high-throughput simulations, while FAIR-compliant data practices promote interoperability and reuse. The infrastructure also incorporates rich, multimodal datasets, including those derived from microscopy, spectroscopy, and diffraction, providing a holistic environment for materials research.

With thousands of users, millions of dataset downloads, and expanding adoption in academic, industrial, and governmental settings, JARVIS has emerged as a comprehensive infrastructure that accelerates materials innovation through automation, openness, and AI-driven insights. This article provides an overview of selected components, impacts, and community engagement within the JARVIS ecosystem. For further details, we refer readers to our prior publications \cite{choudhary2020joint,wines2023recent}.

\section{Overview of the JARVIS Infrastructure}

\begin{figure}[h]
    \centering
    \includegraphics[trim={0. 0cm 0 0cm},clip,width=1.0\textwidth]{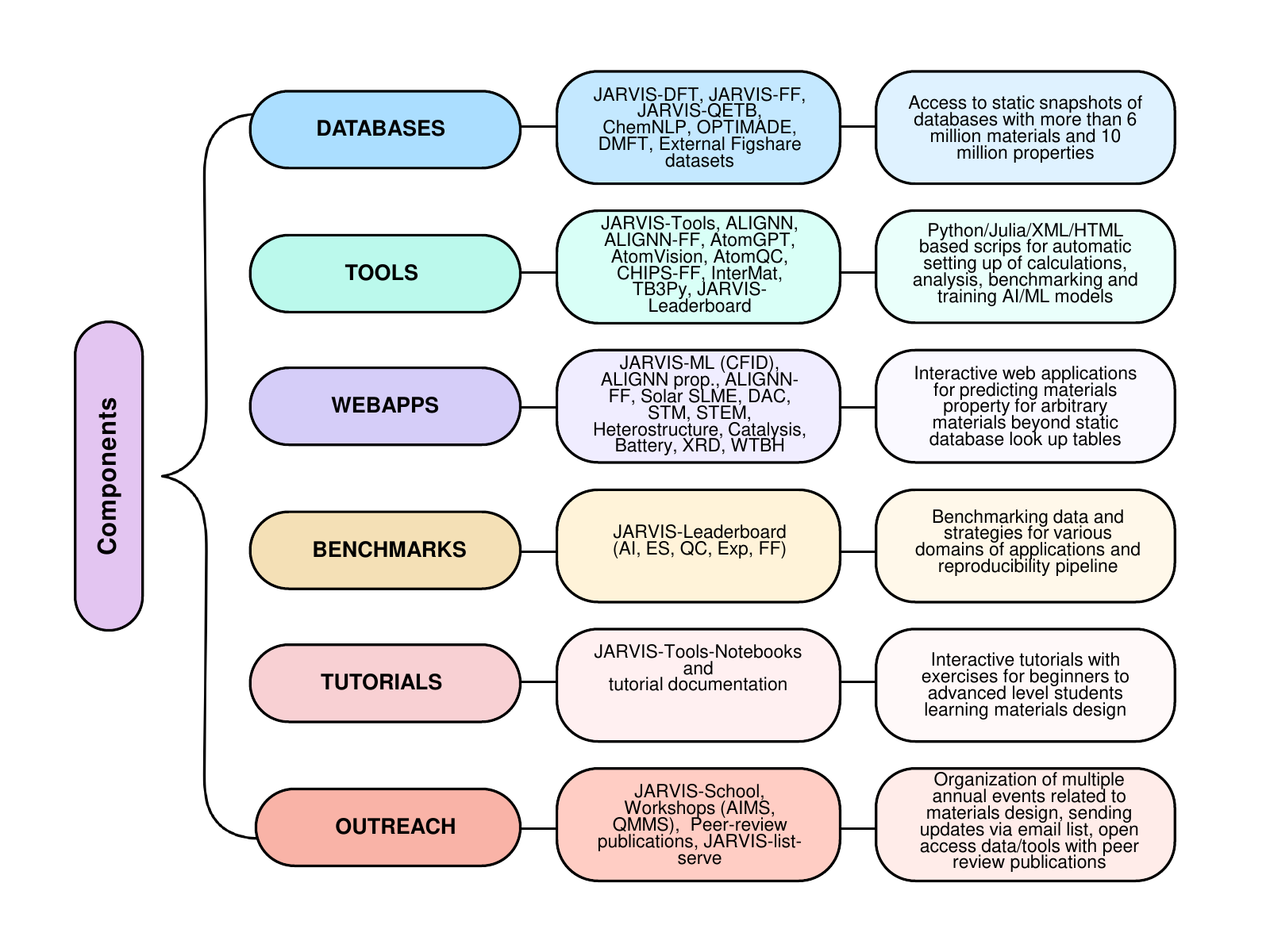}
    \caption{Core Components of the NIST JARVIS Infrastructure. This schematic highlights key elements of JARVIS, including databases for structural, electronic, mechanical, and topological properties, tools for AI/ML modeling, simulations, and benchmarking, tutorials for educational outreach, benchmarking standards for validation against experiments, and community engagement through workshops and collaborations, fostering reproducible materials research. }
    \label{core_components}
\end{figure}
\subsection{Core Components}

JARVIS is a comprehensive infrastructure to accelerate materials discovery and design through the integration of advanced computational tools and extensive datasets. At its core, JARVIS encompasses several key components: 1) databases, 2) software tools, 3) interactive webapps, 4) a large number of benchmarks, 5) step by step tutorials and 6) various events to promote materials design knowledge among students and researchers as shown in Fig. \ref{core_components}. A brief description of each component is given below.

\subsubsection{Databases}

JARVIS hosts a suite of materials databases that provide detailed information on various material properties. These databases consist of its own resources such as JARVIS-DFT as well as external datasets such as Alexandria DB \cite{schmidt2024improving}  in a uniform format to allow seamless integration as well as comparative analysis.  The aggregate of such databases can be up to 6 million materials and 10 million properties. These databases have been downloaded close to 2 million times indicating a wide adaptability by the materials research community. Note that there could be overlap among various databases. To name a few of these, the JARVIS-DFT database contains density functional theory calculations for over 90,000 materials, offering data on structural, electronic, optical, mechanical, phonon, and topological properties. Additionally, the JARVIS-FF database includes classical force-field calculations for approximately 2,000 materials, facilitating the study of interatomic interactions and molecular dynamics simulations. These databases are continuously updated to incorporate new materials and properties, ensuring a comprehensive and current resource for researchers. A list of databases are available at: \url{https://atomgptlab.github.io/jarvis-tools/databases/}. While these tutorials are static snapshots of datasets, a much more user friendly and interactive list of databases is available at: \url{https://jarvis.nist.gov/#Databases}.

\subsubsection{Tools} To support materials design, JARVIS provides a range of software tools that allow setting up and analysis of materials design tasks, leveraging artificial intelligence (AI) and machine learning (ML) models, as well as simulation frameworks. For instance, JARVIS-Tools allow integrating with more than 15 software types such as Vienna Ab initio Simulation Package (VASP) \cite{kresse1996efficient,kresse1996efficiency}, Quantum Espresso (QE) \cite{giannozzi2009quantum}, Large-scale Atomic/Molecular Massively Parallel Simulator (LAMMPS) \cite{plimpton1995fast}, etc., automatic curation and analysis of results, converting information to webpages, providing links to databases mentioned above etc. Similarly, the Atomistic Line Graph Neural Network (ALIGNN) property predictor \cite{choudhary2021atomistic} utilizes graph neural networks to rapidly predict material properties, while the ALIGNN Force-field tool \cite{choudhary2023unified} enables fast structure optimization. All of these tools are open source, available on GitHub and with reasonable documentation to allow both new comers as well as experienced users to utilize resources available in JARVIS. There are hundreds of GitHub stars, and thousands or millions of PyPi downloads of these software tools indicating profound utilization of these novel resources. These tools are designed to streamline the materials design process, thereby reducing the reliance on time-consuming methods such as first-principles calculations.  Currently, there are at least 13 such tools available through JARVIS \textcolor{black}{as shown in Table. 1.}

\begin{table}[h!]
\centering
\caption{JARVIS-related repositories and toolkits for materials design under the GitHub organization: https://github.com/atomgptlab .}
\begin{tabular}{|l|p{10cm}|}
\hline

\textbf{Name} & \textbf{Short Description} \\
\hline
\texttt{1.jarvis-tools} & JARVIS-Tools: An open-source software package for data-driven atomistic materials design \\
\hline
\texttt{2.alignn} & ALIGNN: Atomistic Line Graph Neural Network and force-field \\
\hline
\texttt{3.atomgpt} & AtomGPT: Atomistic Generative Pretrained Transformer for Forward and Inverse Materials Design \\
\hline
\texttt{4.chemnlp} & ChemNLP: A Natural Language Processing based Library for Materials Chemistry Text Data \\
\hline
\texttt{5.atomvision} & AtomVision: Deep learning framework for atomistic image data \\
\hline
\texttt{6.atomqc} & AtomQC: Atomistic Calculations on Quantum Computers \\
\hline
\texttt{7.tb3py} & TB3Py: Two- and three-body tight-binding calculations for materials \\
\hline
\texttt{8.intermat} & InterMat: Interface materials design toolkit \\
\hline
\texttt{9.defectmat} & DefectMat: Defect materials design toolkit \\
\hline
\texttt{10.catalysismat} & CatalysisMat: Catalytic materials design toolkit \\
\hline
\texttt{11.jarvis-tools-notebooks} & A Google-Colab Notebook Collection for Materials Design \\
\hline

\texttt{12.jarvis\_leaderboard} & JARVIS-Leaderboard: Explore State-of-the-Art Materials Design Methods and Reproducible Benchmarks \\
\hline
\hline

\end{tabular}
\end{table}

\subsubsection{Tutorials} Recognizing the importance of education and accessibility, JARVIS offers a collection of Jupyter, Google Colab and Serverless Materials Design (SLMat) \cite{choudhary2024slmat} notebooks that serve as tutorials for various methods in materials design. These resources cover topics such as electronic structure calculations, force-field development, AI applications, and quantum computing methods. By providing these educational materials, JARVIS aims to equip researchers and students with the necessary skills to effectively utilize its tools and databases in their work. One of the major components of tutorials is the JARVIS-Tools-Notebooks framework \url{https://github.com/atomgptlab/jarvis-tools-notebooks} available on GitHub. Currently, there are more than 100 notebooks available in this repository. Another set of documented tutorials are available at \url{https://atomgptlab.github.io/jarvis-tools/tutorials/} .

\subsubsection{Benchmarking Standards}

Rigorous, transparent benchmarking is essential for reproducible materials research.  
To this end, JARVIS hosts the JARVIS‑Leaderboard (\url{https://pages.nist.gov/jarvis_leaderboard/}), an open-source, community‑driven platform that evaluates methods spanning the full materials‐design workflow~\cite{choudhary2024jarvis}.  The leaderboard is built with \texttt{mkdocs}, version‑controlled on GitHub, and rebuilt automatically via continuous‑integration (CI) tests, ensuring that every contribution can be reproduced from code, data and metadata.

At the time of writing, the leaderboard contains 322 benchmarks, 2087 contributions, and more than 8.75 million individual data points drawn from 503 distinct methods and maintained by 26 active contributors. Benchmarks are organised into five top‑level categories-Artificial Intelligence (AI), Electronic Structure (ES), Force-Fields (FF), Quantum Computation (QC) and Experiments (EXP) with contributions currently distributed in following numbers: AI 1034, ES 741, FF 282, QC 6 and EXP 25.

Each task is defined by a machine‑readable folder that bundles: (i) a ground‑truth dataset (JSON/CSV, optionally zipped), (ii) one or more evaluation metrics (e.g.\ MAE, RMSE, accuracy, MULTIMAE/L1 norm for spectra), and (iii) minimal metadata (\texttt{metadata.json}) describing the target property, units, data split, licence and citation.   
The detailed contributor guide (\url{https://pages.nist.gov/jarvis_leaderboard/guide/guide/}) walks users through forking the GitHub repository, formatting predictions as \texttt{*.csv.zip}, adding a \texttt{run.sh}/\texttt{Dockerfile} for full reproducibility, and submitting a pull‑request that triggers CI tests and eventually publishes results on the static. At present, the primary goal of the JARVIS-Leaderboard is to provide a consistent and reproducible benchmarking framework for comparing models and methods across multiple material properties. While it does not currently track historical changes in property predictions for individual compounds, this is a valuable direction for future development.

\subsubsection{Outreach and Community Engagement}

JARVIS is committed to fostering a collaborative scientific community through various outreach initiatives. The platform hosts workshops, webinars, and collaborative projects to engage with researchers, educators, and industry professionals. These efforts aim to disseminate knowledge, gather feedback, and promote the adoption of JARVIS tools and databases in the broader materials science community. Some of these events include JARVIS-School, Artificial Intelligence for Materials Science (AIMS) and Quantum Matters in Material Sciences (QMMS) workshops. Currently, we have hosted more than 4 AIMS, 3 QMMS and 10 JARVIS-Schools. Most of the video recordings are available on our website \url{https://jarvis.nist.gov/events/}.

Collectively, these components establish JARVIS as a robust and versatile infrastructure that supports the entire materials design lifecycle, from data generation and analysis to education and community engagement. 

\subsection{Adhering to FAIR Principles}
The JARVIS infrastructure is committed to the FAIR principles-Findability, Accessibility, Interoperability, and Reproducibility-to enhance data management and utilization in materials science.

\subsubsection{Findability} Findability is achieved through comprehensive metadata and indexing, enabling users to efficiently locate datasets and tools within the JARVIS platform. The infrastructure's databases, such as JARVIS-DFT, are meticulously organized, allowing researchers to search for materials based on various properties and criteria. 

\subsubsection{Accessibility} Accessibility is ensured by providing open access to JARVIS datasets and tools. Researchers can freely retrieve and utilize data without restrictive barriers, promoting widespread use and collaboration. The platform's user-friendly web interface and Application Programming Interface (API)s facilitate seamless data access and integration into external workflows. 

\subsubsection{Interoperability} Interoperability is supported through the use of standardized data formats and protocols for datasets, analysis tools, web design, benchmarking procedures, enabling integration with other databases and computational tools. JARVIS employs common file formats and adheres to community standards, ensuring compatibility and facilitating data exchange across various platforms. 

\subsubsection{Reusability} Reusability is a cornerstone of JARVIS, achieved by providing detailed documentation, workflows, and tutorials. The platform offers Jupyter and Google Colab notebooks that guide users through data analysis and simulation processes, ensuring that studies can be replicated and validated by others. 

By adhering to these FAIR principles, JARVIS enhances the efficiency, transparency, and collaborative potential of materials research, fostering an environment where data and tools are readily available and usable by the global scientific community.

\section{Theoretical and Experimental Method Integration}

\begin{figure}[hbt!]
    \centering
    \includegraphics[trim={0. 0cm 0 0cm},clip,width=1.0\textwidth]{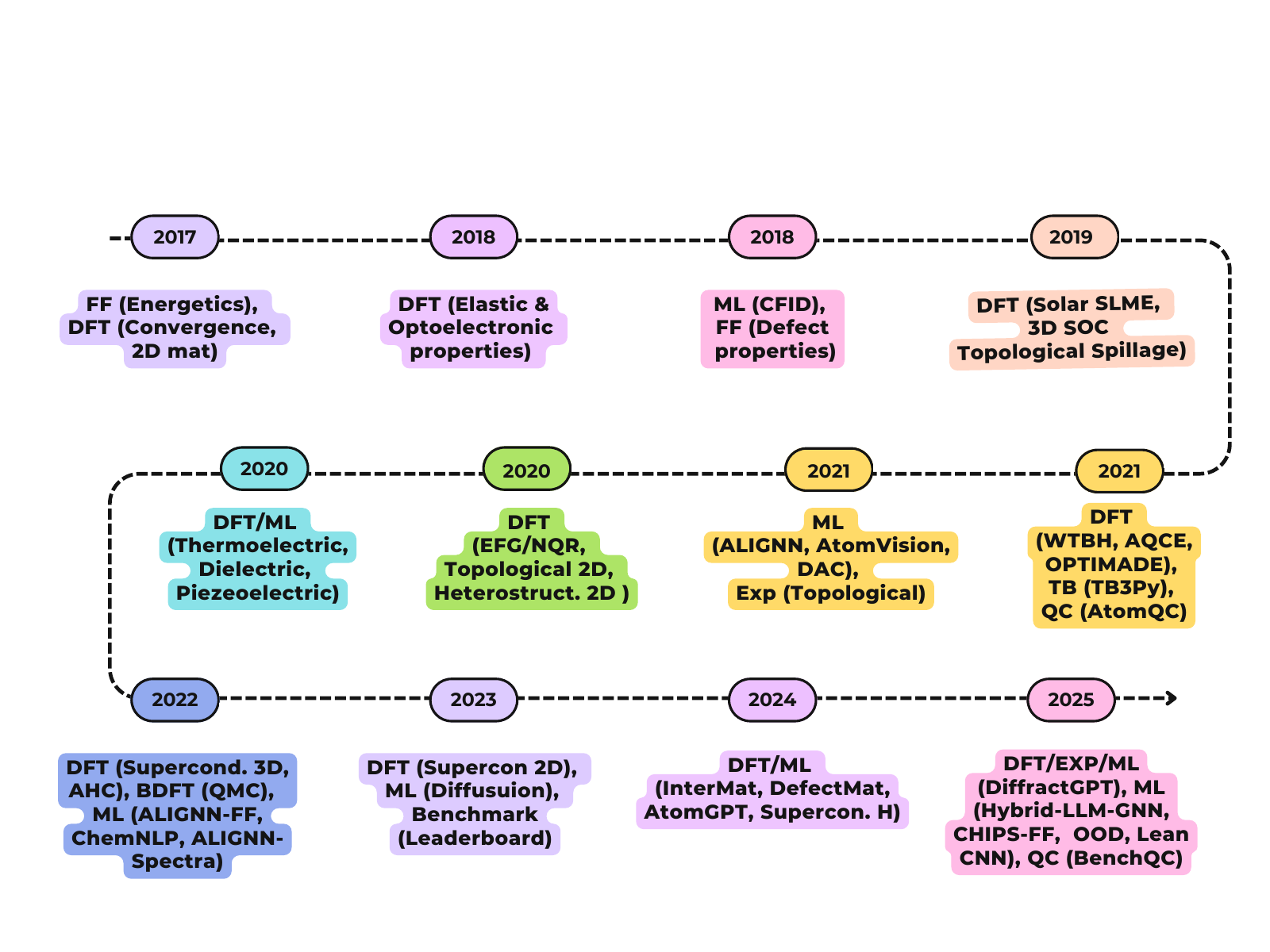}
    \caption{Evolution of Different Projects in the NIST JARVIS Infrastructure (2017-2025). This figure illustrates the timeline and development of various JARVIS initiatives, including JARVIS-DFT, JARVIS-FF, JARVIS-ML, JARVIS-Exp, and other related projects within the NIST JARVIS ecosystem. The steady expansion from 2017 to 2025 highlights the integration of density functional theory (DFT) databases, machine learning models, force field developments, and experimental data repositories. The evolution reflects increasing computational capabilities, data-driven approaches, and interdisciplinary collaborations aimed at advancing materials design and discovery. }
    \label{timeline}
\end{figure}
Several theoretical and experimental methods have been developed, utilized, and integrated within JARVIS over the course of 7+ years as shown in Fig. \ref{timeline}. It shows the timeline and development of various JARVIS initiatives, including JARVIS-DFT, JARVIS-FF, JARVIS-ML, JARVIS-Exp, ALIGNN, Atomistic Generative Pretrained Transformer (AtomGPT) \cite{choudhary2024atomgpt} and other related projects within the JARVIS ecosystem. The steady expansion from 2017 to 2025 highlights the integration of density functional theory (DFT) databases, machine learning models, force field developments, and experimental data repositories. The evolution reflects increasing computational capabilities, data-driven approaches, and interdisciplinary collaborations aimed at advancing materials design and discovery. We can categorize these methods into quantum mechanical, machine learning, classical and experimental categories. A brief description of these components is given below.

\subsection{Quantum Mechanical Methods}

\begin{table}[htb!]
\centering
\begin{tabular}{|c|c|c|c|}
\hline
& \textbf{JARVIS-DFT} & \textbf{MP} & \textbf{OQMD} \\
\hline
\hline
Materials & 90000 & 144595 & 1022663 \\
\hline
DFT methods &OptB88vdW,mBJ,SOC &PBE,SCAN & PBE \\
\hline
K-point/cut-off & Converged per mat. & Fixed  & Fixed\\
\hline
SCF convergence criteria & Energy \& Forces & Energy & Energy \\
\hline
Elastic tensors/phonons & 17402 & 14072 & - \\
\hline
Piezoelectric, IR spect. & 4801 & 3402 & - \\
\hline
Dielectric tensors (w/o ion) & 4801 (15860) & 3402 & - \\
\hline
Electric field gradients & 11865 & - & - \\
\hline
XANES spectra & - & 22000 & - \\
\hline
2D monolayers & 1011 & - & - \\
\hline
Raman spectra & 400 & 50 & - \\
\hline
Seebeck, Power F & 23210 & 48000 & - \\
\hline
Solar SLME & 8614 & - & - \\
\hline
SOC Spillage & 11383 & - & - \\
\hline
WannierTB & 1771 & - & - \\
\hline
STM images & 1432 & - & - \\
\hline
Supercon Tc & 2200 & - & - \\
\hline
Vacancy & 400/192494 & - & - \\
\hline
Surfaces & 300 & - & - \\
\hline
Interfaces & 600/1.4 trill. & - & - \\
\hline
\end{tabular}
\caption{\textcolor{black}{Comparison of three major computational DFT-based materials databases: JARVIS-DFT, Materials Project (MP), and the Open Quantum Materials Database (OQMD). The table summarizes key characteristics and property coverage across the platforms, highlighting both general metrics (e.g., number of materials and DFT methods) and specific physical properties computed. }}
\label{tab:materials_databases}
\end{table}

The JARVIS infrastructure integrates a range of quantum mechanical methods to provide comprehensive insights into material properties. 
\subsubsection{Density Functional Theory and Tight-binding }

Because of its ab initio nature and wide applicability, Density Functional Theory (DFT) serves as a cornerstone of the JARVIS infrastructure. DFT is a quantum mechanical modeling method used to investigate the electronic structure of many-body systems, particularly in condensed matter physics and materials science. The JARVIS-DFT database provides a comprehensive repository of computed structural, electronic, optical, and mechanical properties for a wide array of materials \cite{choudhary2018elastic,choudhary2018computational,choudhary2019accelerated,choudhary2020joint}, making it a valuable resource for the materials research community.

JARVIS includes two primary DFT-based datasets: JARVIS-DFT, primarily built using the Vienna Ab-initio Simulation Package (VASP), and JARVIS-QETB, based on Quantum ESPRESSO calculations. These databases span a range of DFT fidelities, including local density approximation (LDA), generalized gradient approximation (GGA), meta-GGA, van der Waals (vdW)-corrected functionals such as OptB88vdW, and beyond, including many-body perturbation theory methods. The JARVIS-Leaderboard framework further standardizes these calculations for benchmarking and model validation.

The JARVIS-QETB database was also instrumental in the development of a universal tight-binding model across the periodic table \cite{garrity2023fast}. This model includes two-body and three-body effective interaction terms along with self-consistent charge transfer, allowing it to accurately describe metallic, covalent, and ionic bonds with a unified parameter set.

\textcolor{black}{To contextualize the role of JARVIS among other leading DFT databases, Table~\ref{tab:materials_databases} provides a comparison of three prominent platforms: JARVIS-DFT, the Materials Project (MP), and the Open Quantum Materials Database (OQMD). The table summarizes key characteristics and the breadth of physical properties computed across platforms. JARVIS-DFT emphasizes per-material convergence of k-point meshes and plane-wave cutoffs, and includes a number of uniquely reported quantities such as electric field gradients, spin-orbit coupling (SOC) spillage, two-dimensional monolayers, scanning tunneling microscopy (STM) images, and superconducting transition temperatures. In contrast, MP features a broad range of high-throughput properties including XANES spectra and thermoelectric transport coefficients, while OQMD prioritizes scale, offering formation energies for over one million materials. Numbers in parentheses in the table indicate data overlaps, such as the 41,697 structures common to both JARVIS and MP. The vacancy and interface rows denote the number of computed entries and machine learning model based predictions (e.g., 400 vacancy defects with DFT and 192,494 ALIGNN based predictions; 600 computed interfaces with alternating slab junction model using DFT and 1.4 trillion interfaces analyzed with ALIGNN). Together, these databases exhibit complementary strengths: JARVIS prioritizes depth and diversity of physical property predictions, MP provides extensive high-throughput coverage, and OQMD offers a broad-scale energetics dataset. Note that all platforms are continuously evolving, and database sizes are subject to change.}

\subsubsection{Dynamical Mean Field Theory}
In addition to DFT, JARVIS incorporates more advanced quantum mechanical approaches to address the limitations of standard DFT, particularly in dealing with strongly correlated electron systems \cite{mandal2019systematic,choudhary2021quantum}. One such method is Dynamical Mean Field Theory (DMFT), which provides a non-perturbative treatment of local interactions and is effective in capturing the physics of correlated materials. By integrating DMFT, JARVIS enhances its capability to predict electronic properties in materials where electron correlation plays a significant role. 

\subsubsection{Quantum Monte Carlo}
Furthermore, JARVIS employs Quantum Monte Carlo (QMC) \cite {wines2023systematic,wines2024combined} methods, which are stochastic approaches used to solve the Schrödinger equation with high accuracy. QMC is particularly useful for benchmarking DFT results and providing insights into systems where DFT may be inadequate. The inclusion of QMC calculations within JARVIS allows for more reliable predictions of material properties, especially in cases involving complex electronic interactions. 

\subsubsection{Quantum Computing}
Recognizing the emerging potential of quantum computing in materials science, JARVIS also explores quantum computing algorithms for simulating material properties \cite{pollard2025benchqc,choudhary2021quantum}. By interfacing with quantum computing frameworks such as Qiskit, JARVIS aims to leverage quantum algorithms to solve problems that are computationally intensive for classical computers, such as the simulation of strongly correlated systems and the calculation of electronic spectra. This integration positions JARVIS at the forefront of incorporating cutting-edge computational techniques in materials research. 

By encompassing these diverse quantum mechanical methods, JARVIS provides a robust and versatile platform for the accurate prediction and analysis of material properties, catering to a wide range of research needs in the field of materials science.

\subsection{Machine Learning and Data-Driven Approaches}
\subsubsection{Fingerprinting Techniques}
The JARVIS infrastructure leverages advanced machine learning and data-driven approaches to enhance materials discovery and design. A key aspect of this strategy involves the use of fingerprinting techniques, which generate unique representations of materials' structural and compositional features such as element fraction descriptors, average chemical descriptors and classical force-field inspired descriptors (CFID) descriptors \cite{choudhary2018machine}. JARVIS is also integrated within other fingerprinting tools such as MatMiner \cite{ward2018matminer} to allow various other schemes of fingerprints which can then be compared for instance in the JARVIS-Leaderboard \cite{choudhary2024jarvis} for accuracy and speed. These fingerprints serve as inputs for machine learning models. These models have recently been analyzed for their uncertainty quantification applications as well \cite{tavazza2021uncertainty}. Very often, uncertainty quantification metrics can be more important than typical machine learning model metrics while implementing realistic materials design.

\subsubsection{Graph Neural Networks}
JARVIS employs Graph Neural Networks (GNNs) to model the complex relationships within materials. Such GNNs have been used for atomic structure and atomistic image analysis in ALIGNN \cite{choudhary2021atomistic}, ALIGNN-FF \cite{choudhary2023unified} and AtomVision \cite{choudhary2023atomvision} frameworks. They are becoming an inevitable part of almost all materials design processes. GNNs are adept at handling graph-structured data, making them well-suited for representing materials where atoms are nodes and bonds are edges. The Atomistic Line Graph Neural Network (ALIGNN) is a notable example, which enhances traditional GNNs by incorporating both bond lengths and bond angles into its architecture. This inclusion of angular information has led to improved performance in predicting a wide range of material properties, as demonstrated in studies utilizing the JARVIS-DFT and other external datasets.  

\subsubsection{Transformer Models}
In addition to GNNs, JARVIS explores the application of transformer-based models in materials science. Originally developed for natural language processing( NLP such as in ChemNLP \cite{choudhary2023chemnlp}), Transformers have been adapted to handle the complex patterns found in material structures for both forward and inverse materials design such as in AtomGPT \cite{choudhary2024atomgpt}, DiffractGPT \cite{choudhary2024diffractgpt} and MicroscopyGPT \cite{choudhary2025microscopygpt} frameworks. Recent research has introduced models that generate atomic embeddings using Transformer architectures, leading to enhanced predictions of material properties. These models capture long-range dependencies and intricate relationships within the material data, offering a powerful tool for materials informatics. 

By integrating these machine learning methodologies-fingerprinting techniques, Graph Neural Networks, and Transformer-based models-JARVIS provides a robust framework for the accelerated discovery and design of new materials. 

\subsubsection{Conventional Force Fields}

A fundamental component of JARVIS is Classical Molecular Dynamics (MD), which utilizes empirical interatomic potentials to simulate the behavior of atoms and molecules over time. This method is particularly effective for studying the thermodynamic and kinetic properties of materials at finite temperatures and for modeling large systems that are computationally prohibitive for quantum mechanical methods.

JARVIS-FF~\cite{choudhary2018high,choudhary2017evaluation} maintains a curated repository of classical force fields tailored for various materials, enabling accurate MD simulations with software packages such as the Large-scale Atomic/Molecular Massively Parallel Simulator (LAMMPS)~\cite{plimpton1995fast}, General Utility Lattice Program (GULP)~\cite{gale2003general}, and AMBER~\cite{salomon2013overview}. In its early development, JARVIS-FF built upon foundational work by Chandler Becker~\cite{becker2013considerations} and leveraged contributions from related repositories such as OpenKIM~\cite{waters2023automated}.

\textcolor{black}{It is often difficult to determine \emph{a priori} whether a given force field is suitable for a particular material or property, especially when ground truth data are unavailable or inconsistent. JARVIS-FF addresses this challenge by linking high-throughput FF calculations directly to high-fidelity DFT reference data in the JARVIS-DFT database, enabling 1:1 benchmarking across properties such as lattice constants, formation energies, phonon spectra, and elastic moduli.} \textcolor{black}{This benchmarking-oriented approach goes beyond merely hosting force fields by providing automated workflows (e.g., SLURM-compatible job scripts), ready-to-use LAMMPS input templates, and comprehensive metadata tracking. All simulation scripts and examples, including Google Colab notebooks, are openly accessible at \url{https://github.com/JARVIS-Materials-Design/jarvis-tools-notebooks}.} \textcolor{black}{This infrastructure empowers users to perform reproducible, scalable, and validated MD simulations tailored to their material systems of interest.}

\subsubsection{Machine Learning Force Fields}
To enhance the accuracy and transferability of force fields, JARVIS incorporates Machine Learning Force Fields (MLFFs) \cite{jacobs2025practical} based on ALIGNN. The ALIGNN-FF is seamlessly integrated with an atomic simulation environment (ASE) Calculator to leverage various dynamic tools. These MLFFs are trained on extensive datasets derived from high-fidelity quantum mechanical calculations, enabling them to capture complex potential energy surfaces with high precision. By leveraging machine learning techniques, JARVIS's MLFFs can predict interatomic forces and energies more efficiently than traditional empirical potentials, thereby expanding the scope and scale of simulations that can be performed. This approach allows for rapid and accurate predictions of material properties, facilitating accelerated materials discovery and design. 

By incorporating classical MD and MLFFs, JARVIS offers a robust and flexible infrastructure for the multiscale simulation of materials, accommodating a wide range of systems and properties with varying degrees of complexity.

\subsection{Experimental Data Integration}
The JARVIS infrastructure emphasizes the integration of experimental data to validate and enhance its theoretical predictions, thereby bridging the gap between computation and real-world observations. A key aspect of this integration involves the incorporation of experimental data for benchmarking materials properties predicted with quantum/classical methods. While most of the benchmarking experimental data comes from previous experiment literature, JARVIS has its own experimental data including microscopy, diffraction and cryogenics experiments. By comparing simulated data with experimental data, JARVIS ensures the accuracy and reliability of its computational models. Furthermore, JARVIS integrates data for inorganic, organic and metal-organic compounds. Some examples include experimental measurements of superconductors \cite{wines2023high}, nanoparticles \cite{choudhary2018high}, topological magnets \cite{choudhary2021high}, solar cells \cite{choudhary2018computational}
,  metal organic frameworks \cite{choudhary2022graph} etc. This integration allows for the validation of theoretical predictions against experimentally determined crystal structures, enhancing the robustness of the simulations. 

By correlating theoretical predictions with experimental observations, JARVIS not only validates its computational approaches but also refines its models to better reflect real-world behaviors. This synergy between theory and experiment is crucial for advancing materials science, as it ensures that computational discoveries are grounded in empirical evidence, thereby accelerating the development of new materials with desired properties. We note that while automating theoretical materials design methods are well developed the same for experiments is still lacking. In the near future, JARVIS will focus on automating and benchmarking every distinct aspect of materials experimental methods, enabling truly autonomous self-driving lab experiments and advancing AI-driven materials design.

\section{Coverage Across Material Classes}
\begin{figure}[H]
    \centering
    \includegraphics[trim={0. 0cm 0 0cm},clip,width=1.0\textwidth]{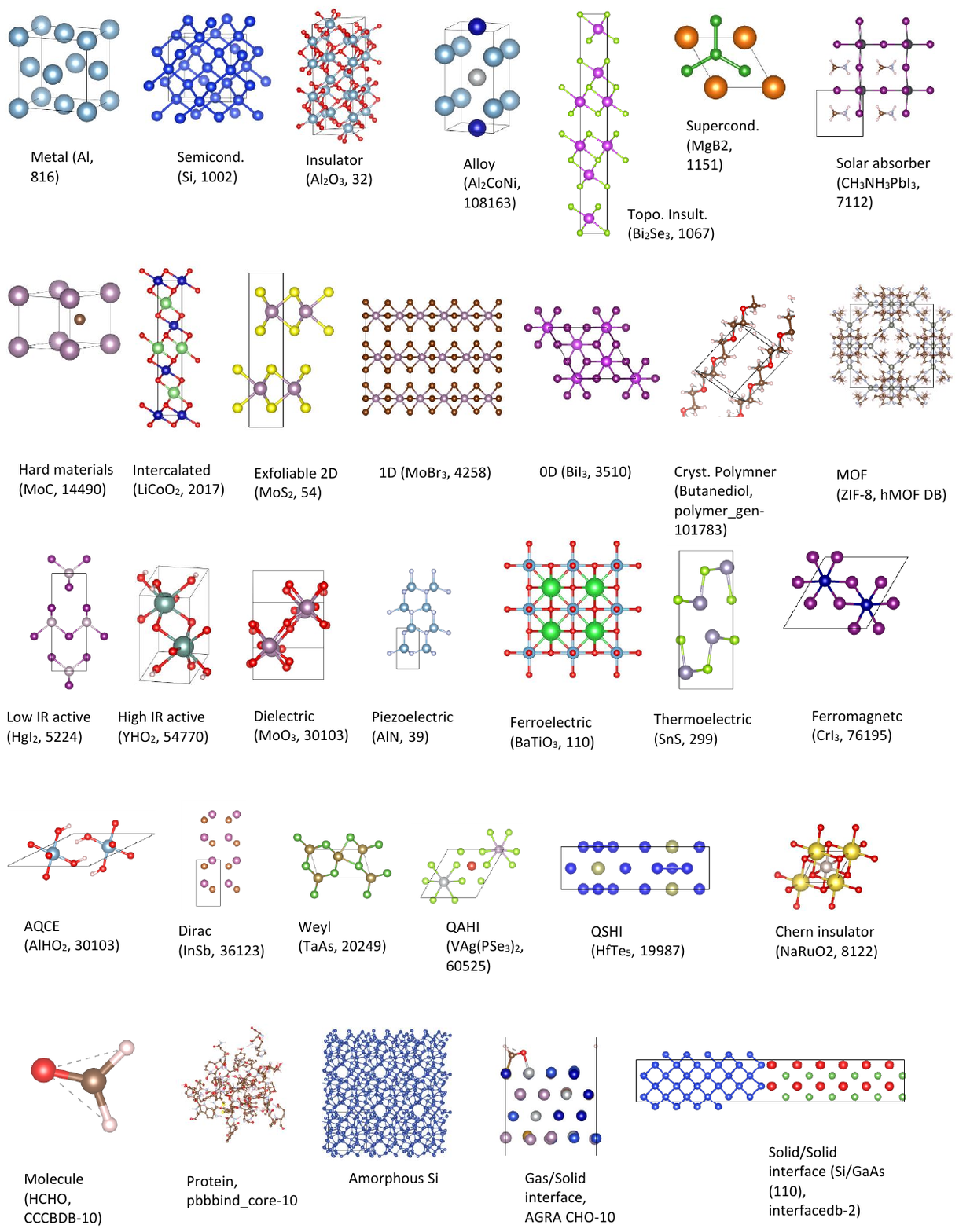}
    \caption{Atomic structure, identifier and chemical formula for various materials classes are provided wherever applicable. Examples include metals, semiconductors, insulators, alloys, topological insulators, superconductors, solar absorbers, mechanically hard materials, intercalated materials for battery cathodes, mechanically exfoliable 2D, 1D, 0D materials, crystalline polymers, metal organic frameworks, low infrared active mode, high infrared active mode materials, materials with dielectric, piezoelectric constants, ferroelectrics, thermoelectrics, 2D ferromagnets, anomalous quantum confinement effect materials (AQCE), Dirac and Weyl semimetals, quantum anomalous hall (QAHI) and spin hall insulators (QSHI), Chern insulators, molecules in CCCBDB database, proteins in external PDB datasets, amorphous materials generated with ALIGNN-FF/DFT, gas/solid interface for catalysis, solid-solid interfaces for microelectronics and other defect based materials.}
    \label{coverage}
\end{figure}

The JARVIS infrastructure offers extensive coverage across a diverse array of material classes, facilitating advancements in both theoretical understanding and practical applications as shown in Fig. \ref{coverage}. It shows various types of materials classes with an example of each available in JARVIS such as metals, semiconductors, insulators, alloys \cite{choudhary2018computational}, topological insulators \cite{choudhary2019high,choudhary2020computational}, superconductors \cite{choudhary2022designing,wines2023high,wines2024data,wines2023inverse}, solar absorbers \cite{choudhary2019accelerated}, mechanically hard materials \cite{choudhary2018elastic}, intercalated materials for battery cathodes, mechanically exfoliable 2D, 1D, 0D materials\cite{choudhary2018elastic,choudhary2017high}, crystalline polymers, metal organic frameworks \cite{choudhary2022graph}, low infrared active mode, high infrared active mode materials, materials with dielectric, piezoelectric constants, ferroelectrics \cite{choudhary2020high}, thermoelectrics \cite{choudhary2020data}, 2D ferromagnets \cite{wines2023systematic}, anomalous quantum confinement effect materials, Dirac and Weyl semimetals, quantum anomalous hall and spin hall insulators \cite{choudhary2020computational}, Chern insulators \cite{choudhary2021high}, molecules in Computational Chemistry Comparison and Benchmark DataBase (CCCBDB) database \cite{johnson1999nist}, proteins in external protein databank (PDB) datasets \cite{liu2015pdb}, amorphous materials generated with ALIGNN-FF/DFT \cite{wines2024chips}, gas/solid interface for catalysis \cite{wang2024examining}, solid-solid interfaces for microelectronics \cite{choudhary2024intermat} and other defect based materials \cite{choudhary2023can}. Unless specified with the database name, the identifier should indicate JARVIS-DFT identifiers. A brief description of some of these classes is given below.
\newpage
\subsection{Metals, Semiconductors, and Insulators}
Based on electronic bandgaps, materials can be classified into metal, semiconductor and insulator categories. JARVIS provides comprehensive datasets on these fundamental material categories, encompassing detailed information on their structural, electronic, and mechanical properties. This data is crucial for applications ranging from electronic device fabrication to structural engineering. For instance, the platform's resources aid in the selection and optimization of materials for semiconductors, which are integral to modern electronics. 

\subsection{Superconductors and Quantum Materials}

All materials are quantum in nature, but when these effects show dominance at the classical level as well, then they can be utilized for numerous technological applications. Recognizing the significance of quantum materials in advancing technology, JARVIS includes data on superconductors and other quantum materials such as topological insulators, anomalous Hall insulators, Dirac and Weyl semimetals etc. This encompasses properties such as critical temperatures and electronic band structures, which are vital for developing applications like quantum computing and highly efficient power transmission systems. 

\subsection{Carbon Capture and Sustainable Materials}

Addressing environmental challenges, JARVIS integrates data on materials designed for carbon capture and sustainability. This includes information on metal-organic frameworks (MOFs), zeolites, polymeric membranes and other porous materials that can adsorb carbon dioxide, contributing to efforts in mitigating climate change. 

\subsection{High-Strength and Structural Materials}

For applications requiring materials with exceptional strength and durability, such as in the construction and aerospace industries, JARVIS can be used to rank the strength of materials based on the elastic modulus and derived data. This information assists in the design and selection of materials that meet specific mechanical performance criteria. 

\subsection{Low-Dimensional Materials}

The dimensionality of a material can be defined based on the presence of vdW bonding in 1,2,3 crystallographic directions. JARVIS provides extensive data on 0D, 1D, 2D materials and their properties, which have unique electronic and mechanical properties. This includes information on materials like graphene and transition metal dichalcogenides, which are being explored for next-generation electronic devices and sensors. 

\subsection{Defects and Interfaces}

Understanding defects and interfaces/heterostructures is crucial for tailoring material properties for specific applications\textcolor{black}{ \cite{tran2016surface,zheng2020grain,olmsted2009survey}}. JARVIS includes data on various defect structures and their properties such as vacancies, enabling researchers to predict how these imperfections can influence material behavior. This is particularly important in semiconductors, where defects can significantly impact electronic properties. Similarly, JARVIS provides databases, tools, webapps, and benchmarks on solid-molecule interface (such as for catalysts) and solid-solid interface (such as for heterostructures used in microelectronics) are crucial for materials design.

By encompassing such a wide range of material classes, JARVIS serves as a valuable resource for researchers and engineers, supporting the development of innovative materials and technologies across multiple industries. 

\section{JARVIS as a Platform for Reproducible Science}

The JARVIS serves as a comprehensive platform dedicated to promoting reproducible science in materials research. There are various strategies to enhance reproducibility and descriptions of these strategies are briefly given below. 

\subsection{Peer-Reviewed Publications and Open-Access Data}
JARVIS emphasizes transparency by providing open-access datasets and disseminating findings through peer-reviewed publications. Ideally, each publication in JARVIS should have: 1) fully accessible preprint on arXiv/ChemarXiv etc., 2) code available in GitHub or similar platforms, 3) static data available in Figshare within JARVIS-Tools or similar platforms, 4) benchmarks used for methods available in JARVIS-Leaderboard, 5) WebApp to enhance the applicability of dataset, tools, models generated. This approach ensures that the scientific community can readily access and validate the data, fostering trust and facilitating further research. The platform's commitment to open science is evident in its extensive databases, such as JARVIS-DFT, which offers a wealth of information on material properties. 

\subsection{Interactive Web Applications}
To enhance user engagement and accessibility, JARVIS offers a suite of web applications that provide intuitive interfaces for data retrieval and analysis. These tools enable researchers to explore datasets, perform simulations, and visualize results without the need for extensive computational resources, thereby lowering the barrier to entry for materials research. A list of such apps and brief description are given in Table 3.

\begin{table}[htbp]
\centering
\footnotesize     
\begin{tabular}{|p{3.2cm}|p{5.0cm}|p{7.2cm}|}
\hline
\textbf{Name} & \textbf{URL} & \textbf{Description} \\
\hline
1.ALIGNN Property Predictor & \url{https://jarvis.nist.gov/jalignn/} & Atomistic Line Graph Neural Network (ALIGNN) for rapid property prediction. \\
\hline
2.ALIGNN Force‑Field & \url{https://jarvis.nist.gov/jalignnff/} & GNN‑based interatomic potential for fast structure optimisation. \\
\hline
3.Solar Cells & \url{https://jarvis.nist.gov/jarvissolar} & Estimates theoretical photovoltaic performance of a material. \\
\hline
4.Direct Air Capture & \url{https://jarvis.nist.gov/jdac} & Predicts \(\mathrm{CO_2}\) adsorption isotherms for metal‑organic frameworks. \\
\hline
5.Scanning Tunnelling Microscopy & \url{https://jarvis.nist.gov/jarvisstm} & Generates Tersoff-Hamann STM images from crystal structures. \\
\hline
6.Scanning Transmission Electron Microscopy & \url{https://jarvis.nist.gov/jstem} & Simulates STEM images using the convolution approximation. \\
\hline
7.Heterostructure Builder & \url{https://jarvis.nist.gov/jarvish} & Creates interfaces/heterostructures with the Zur lattice‑matching algorithm. \\
\hline
8.Catalysis (Adsorption) & \url{https://jarvis.nist.gov/jcatalysis} & Predicts adsorption energies of molecules on catalytic substrates. \\
\hline
9.Visualization & \url{https://jarvis.nist.gov/jarvisviz} & Lightweight, in‑browser atomic structure visualiser. \\
\hline
10.JARVIS‑XRD & \url{https://jarvis.nist.gov/jxrd} & Computes theoretical X‑ray diffraction patterns. \\
\hline
11.JARVIS‑WTBH & \url{https://jarvis.nist.gov/jarviswtb} & Calculates properties from Wannier tight‑binding Hamiltonians. \\
\hline
12.JARVIS‑Battery & \url{https://jarvis.nist.gov/jbattery} & Predicts capacities and voltage profiles of battery electrodes. \\
\hline
13.JARVIS‑ML / CFID & \url{https://jarvis.nist.gov/jarvisml} & Property prediction with Classical Force‑Field Inspired Descriptors. \\
\hline
\end{tabular}
\caption{Selected JARVIS web applications and their functionalities in JARVIS.}
\label{tab:jarvis_tools}
\end{table}

\subsection{Scripts and Notebooks for Reproducibility} 

Recognizing the importance of reproducibility, JARVIS provides a collection of scripts and Jupyter notebooks that detail the methodologies used in data generation and analysis. These resources allow researchers to replicate studies, validate findings, and build upon existing work, ensuring that scientific discoveries are robust and verifiable. 

\subsection{Benchmarking and Validation} JARVIS has developed the JARVIS-Leaderboard, an open-source, community-driven platform designed to facilitate benchmarking across various materials design methods as mentioned above. This initiative allows for the comparison of computational predictions with experimental datasets, providing a framework for the performance evaluation of different computational approaches. By systematically validating methods against empirical data, JARVIS ensures the reliability and accuracy of its predictive models. Through these initiatives, JARVIS not only advances materials science but also sets a standard for reproducibility and transparency in computational research. 

\section{Impact and Community Adoption}

\begin{figure}[hbt!]
    \centering
    \includegraphics[trim={0. 0cm 0 0cm},clip,width=1.0\textwidth]{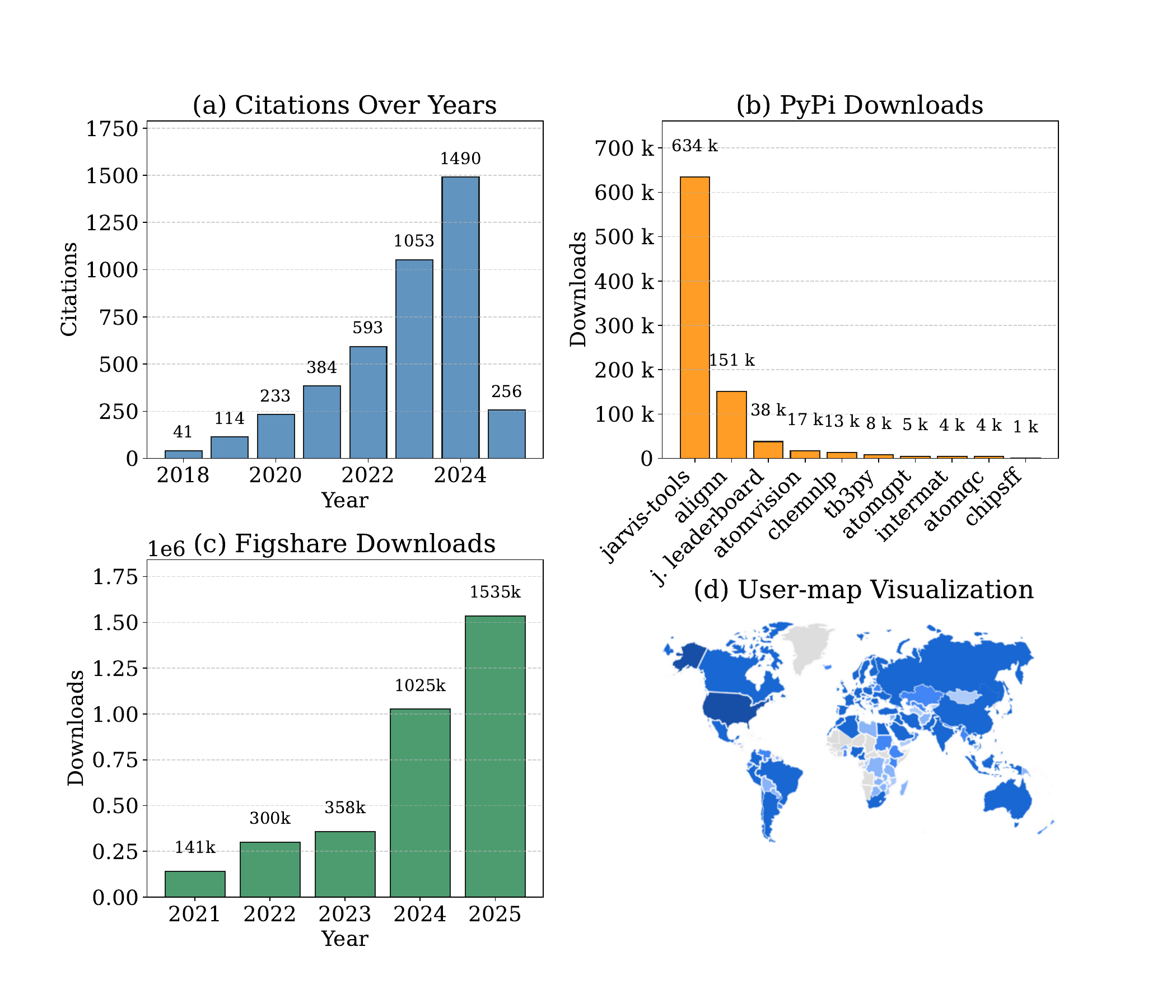}
    \caption{Trends in Citations, Software Downloads, and Geospatial Data Visualization. (a) Annual citation count from 2018 to 2025, shows a significant increase in recent years. (b) PyPi downloads for various materials science and AI-related tools, formatted using engineering notation for clarity. (c) Figshare downloads over time, highlighting adoption trends across different years. Note that some of the PyPI and Figshare download metrics may be influenced by intentional or unintentional automation using bots. Therefore, these numbers should be interpreted with caution. (d) A map visualization provides geospatial insights into usage patterns. Darker blue indicates a higher number, while lighter shades represent lower values.}
    \label{impact}
\end{figure}
The JARVIS has established itself as a pivotal resource in the materials science community, fostering global engagement and collaboration across academia and industry.

\subsection{Global User Base and Downloads} Since its inception in January 2017, JARVIS has garnered a substantial international user base, with over 150,000 users accessing its resources. The papers in JARVIS have been cited over 4000 times. The platform's extensive databases, encompassing more than 90,000 materials and over a million calculated properties, have been downloaded over a million times, underscoring its widespread adoption and utility in the research community. Examples of impact in terms of paper citations, software and data downloads, and global user base are shown in Fig. \ref{impact}.

\subsection{Industry and Academic Collaborations} JARVIS actively supports real-world research and development by facilitating collaborations between academic institutions and industry partners. By providing open-access data and computational tools, JARVIS enables researchers to accelerate materials discovery and optimization, bridging the gap between theoretical research and practical applications. These collaborations have led to advancements in various fields, including electronics, energy storage, and sustainable materials. 

In addition, to promote education and skill development, JARVIS organizes events such as the JARVIS-School, AIMS and QMMS workshops which offer tutorials and hands-on sessions on topics like electronic structure calculations, machine learning applications, and quantum computations as mentioned in the above section of outreach. These programs are designed to introduce participants to open-access databases and tools for materials design, fostering a deeper understanding and practical proficiency in modern computational methods.


\subsection{JARVIS in Education and Curriculum Development}
Beyond workshops, JARVIS aims to contribute to curriculum development by providing educational resources that integrate into academic programs. The platform's collection of Jupyter and Google Colab notebooks serves as valuable teaching aids, allowing students to engage with real-world data and computational tools in a classroom setting. This integration enhances the learning experience and prepares students for careers in materials science and engineering. 

Through these initiatives, JARVIS not only advances materials research but also plays a crucial role in education and the dissemination of knowledge, supporting the development of the next generation of scientists and engineers. 

\section{Future Directions and Open Challenges}
The JARVIS is poised to advance materials science through several strategic initiatives, while also addressing existing challenges.

\subsection{Expanding Materials Coverage} The number of possible materials and types is huge. JARVIS aims to broaden its database in the future by incorporating a wider array of materials, including complex compounds, alloys, and emerging materials like altermagnets, multi-layer heterostructures, etc. This expansion will enhance the platform's utility for diverse research applications. 

\subsection{Enhancing AI/ML Capabilities} The integration of advanced artificial intelligence and machine learning techniques is a priority for JARVIS. By developing more sophisticated models, such as graph neural networks and transformer-based architectures, the platform seeks to improve the accuracy and efficiency of property predictions and materials discovery. 

\subsection{Improving Quantum Computing Integration} Recognizing the potential of quantum computing, JARVIS plans to deepen its integration of quantum algorithms to tackle complex simulations that are challenging for classical computers. This includes exploring quantum algorithms for electronic structure calculations and materials optimization. 

\subsection{Addressing Scalability and High-Performance Computing Needs} With growing length-time scale requirements, and database growth, ensuring scalability and efficient data management becomes crucial. JARVIS is investing in high-performance computing resources and optimizing its computational workflows to handle large-scale simulations and data analyses effectively. 

\subsection{Strengthening Experimental-Theoretical Synergy and Autonomous Experiments} Finally, to bridge the gap between theory and experiment, JARVIS is enhancing its integration of experimental data, such as microscopy and spectroscopy results, with computational predictions. This synergy aims to improve the validation of theoretical models and foster a more comprehensive understanding of material behaviors, which are essential for self-driving labs, autonomous experimentation and robotic agent development.

By pursuing these directions, JARVIS is committed to overcoming current challenges and advancing the field of materials science through innovation and collaboration. 

\section{Conclusions}

The JARVIS has significantly advanced materials science by providing an integrated platform that combines electronic structure calculations, machine learning models, and experimental data. This comprehensive approach has facilitated accelerated materials discovery and design, contributing to numerous peer-reviewed publications and the development of open-access datasets. Looking ahead, JARVIS is poised to play a pivotal role in the next decade of materials discovery. By expanding its databases to include emerging materials and enhancing its machine learning capabilities, JARVIS aims to provide more accurate predictions and insights. The integration of quantum computing algorithms and the development of benchmarking platforms like the JARVIS-Leaderboard further position it as a leader in the field. To fully realize these advancements, collaboration with the broader scientific community is essential. Researchers, educators, and industry professionals are encouraged to engage with JARVIS by utilizing its resources, contributing data, and participating in community-driven initiatives. Such collective efforts will enhance the platform's capabilities and drive innovation in materials science.

\section{Data Availability}
The JARVIS data is available at websites \url{https://jarvis.nist.gov/} and \url{https://atomgpt.org/}.

\section{Acknowledgments}
K.C. thanks the National Institute of Standards and Technology as well as the Johns Hopkins University for funding, computational, and data-management resources. This work was performed with funding from the CHIPS Metrology Program, part of CHIPS for America, National Institute of Standards and Technology, U.S. Department of Commerce. Certain commercial equipment, instruments, software, or materials are identified in this paper in order to specify the experimental procedure adequately. Such identifications are not intended to imply recommendation or endorsement by NIST, nor it is intended to imply that the materials or equipment identified are necessarily the best available for the purpose. 
 \bibliographystyle{elsarticle-num} 
 \bibliography{cas-refs}

\begin{thebibliography}{10}
\expandafter\ifx\csname url\endcsname\relax
  \def\url#1{\texttt{#1}}\fi
\expandafter\ifx\csname urlprefix\endcsname\relax\def\urlprefix{URL }\fi
\expandafter\ifx\csname href\endcsname\relax
  \def\href#1#2{#2} \def\path#1{#1}\fi

\bibitem{callister2020fundamentals}
W.~D. Callister~Jr, D.~G. Rethwisch, Fundamentals of materials science and engineering: an integrated approach, John Wiley \& Sons, 2020.

\bibitem{olson1997computational}
G.~B. Olson, Computational design of hierarchically structured materials, Science 277~(5330) (1997) 1237--1242.

\bibitem{rajan2015materials}
K.~Rajan, Materials informatics: The materials “gene” and big data, Annual Review of Materials Research 45~(1) (2015) 153--169.

\bibitem{jain2013commentary}
A.~Jain, S.~P. Ong, G.~Hautier, W.~Chen, W.~D. Richards, S.~Dacek, S.~Cholia, D.~Gunter, D.~Skinner, G.~Ceder, et~al., Commentary: The materials project: A materials genome approach to accelerating materials innovation, APL materials 1~(1) (2013).

\bibitem{curtarolo2013high}
S.~Curtarolo, G.~L. Hart, M.~B. Nardelli, N.~Mingo, S.~Sanvito, O.~Levy, The high-throughput highway to computational materials design, Nature materials 12~(3) (2013) 191--201.

\bibitem{choudhary2020joint}
K.~Choudhary, K.~F. Garrity, A.~C. Reid, B.~DeCost, A.~J. Biacchi, A.~R. Hight~Walker, Z.~Trautt, J.~Hattrick-Simpers, A.~G. Kusne, A.~Centrone, et~al., The joint automated repository for various integrated simulations (jarvis) for data-driven materials design, npj computational materials 6~(1) (2020) 173.

\bibitem{tadmor2011modeling}
E.~B. Tadmor, R.~E. Miller, Modeling materials: continuum, atomistic and multiscale techniques, Cambridge University Press, 2011.

\bibitem{schmidt2019recent}
J.~Schmidt, M.~R. Marques, S.~Botti, M.~A. Marques, Recent advances and applications of machine learning in solid-state materials science, npj computational materials 5~(1) (2019) 83.

\bibitem{choudhary2022recent}
K.~Choudhary, B.~DeCost, C.~Chen, A.~Jain, F.~Tavazza, R.~Cohn, C.~W. Park, A.~Choudhary, A.~Agrawal, S.~J. Billinge, et~al., Recent advances and applications of deep learning methods in materials science, npj Computational Materials 8~(1) (2022) 59.

\bibitem{vasudevan2019materials}
R.~K. Vasudevan, K.~Choudhary, A.~Mehta, R.~Smith, G.~Kusne, F.~Tavazza, L.~Vlcek, M.~Ziatdinov, S.~V. Kalinin, J.~Hattrick-Simpers, Materials science in the artificial intelligence age: high-throughput library generation, machine learning, and a pathway from correlations to the underpinning physics, MRS communications 9~(3) (2019) 821--838.

\bibitem{schleder2019dft}
G.~R. Schleder, A.~C. Padilha, C.~M. Acosta, M.~Costa, A.~Fazzio, From dft to machine learning: recent approaches to materials science--a review, Journal of Physics: Materials 2~(3) (2019) 032001.

\bibitem{saal2013materials}
J.~E. Saal, S.~Kirklin, M.~Aykol, B.~Meredig, C.~Wolverton, Materials design and discovery with high-throughput density functional theory: the open quantum materials database (oqmd), Jom 65 (2013) 1501--1509.

\bibitem{sanchez2018inverse}
B.~Sanchez-Lengeling, A.~Aspuru-Guzik, Inverse molecular design using machine learning: Generative models for matter engineering, Science 361~(6400) (2018) 360--365.

\bibitem{agrawal2016perspective}
A.~Agrawal, A.~Choudhary, Perspective: Materials informatics and big data: Realization of the “fourth paradigm” of science in materials science, Apl Materials 4~(5) (2016).

\bibitem{zunger2018inverse}
A.~Zunger, Inverse design in search of materials with target functionalities, Nature Reviews Chemistry 2~(4) (2018) 0121.

\bibitem{sholl2022density}
D.~S. Sholl, J.~A. Steckel, Density functional theory: a practical introduction, John Wiley \& Sons, 2022.

\bibitem{plimpton1995fast}
S.~Plimpton, Fast parallel algorithms for short-range molecular dynamics, Journal of computational physics 117~(1) (1995) 1--19.

\bibitem{behler2007generalized}
J.~Behler, M.~Parrinello, Generalized neural-network representation of high-dimensional potential-energy surfaces, Physical review letters 98~(14) (2007) 146401.

\bibitem{zhang2018deep}
L.~Zhang, J.~Han, H.~Wang, R.~Car, W.~E, Deep potential molecular dynamics: a scalable model with the accuracy of quantum mechanics, Physical review letters 120~(14) (2018) 143001.

\bibitem{baker20161}
M.~Baker, 1,500 scientists lift, Nature 533 (2016) 452--454.

\bibitem{lee2023machine}
J.~Lee, D.~Park, M.~Lee, H.~Lee, K.~Park, I.~Lee, S.~Ryu, Machine learning-based inverse design methods considering data characteristics and design space size in materials design and manufacturing: a review, Materials Horizons (2023).

\bibitem{gomez2018automatic}
R.~G{\'o}mez-Bombarelli, J.~N. Wei, D.~Duvenaud, J.~M. Hern{\'a}ndez-Lobato, B.~S{\'a}nchez-Lengeling, D.~Sheberla, J.~Aguilera-Iparraguirre, T.~D. Hirzel, R.~P. Adams, A.~Aspuru-Guzik, Automatic chemical design using a data-driven continuous representation of molecules, ACS central science 4~(2) (2018) 268--276.

\bibitem{zhong2022explainable}
X.~Zhong, B.~Gallagher, S.~Liu, B.~Kailkhura, A.~Hiszpanski, T.~Y.-J. Han, Explainable machine learning in materials science, npj computational materials 8~(1) (2022) 204.

\bibitem{zeni2023mattergen}
C.~Zeni, R.~Pinsler, D.~Z{\"u}gner, A.~Fowler, M.~Horton, X.~Fu, S.~Shysheya, J.~Crabb{\'e}, L.~Sun, J.~Smith, et~al., Mattergen: a generative model for inorganic materials design, arXiv preprint arXiv:2312.03687 (2023).

\bibitem{wilkinson2016fair}
M.~D. Wilkinson, M.~Dumontier, I.~J. Aalbersberg, G.~Appleton, M.~Axton, A.~Baak, N.~Blomberg, J.-W. Boiten, L.~B. da~Silva~Santos, P.~E. Bourne, et~al., The fair guiding principles for scientific data management and stewardship, Scientific data 3~(1) (2016) 1--9.

\bibitem{wines2023recent}
D.~Wines, R.~Gurunathan, K.~F. Garrity, B.~DeCost, A.~J. Biacchi, F.~Tavazza, K.~Choudhary, Recent progress in the jarvis infrastructure for next-generation data-driven materials design, Applied Physics Reviews 10~(4) (2023).

\bibitem{curtarolo2012aflow}
S.~Curtarolo, W.~Setyawan, G.~L. Hart, M.~Jahnatek, R.~V. Chepulskii, R.~H. Taylor, S.~Wang, J.~Xue, K.~Yang, O.~Levy, et~al., Aflow: An automatic framework for high-throughput materials discovery, Computational Materials Science 58 (2012) 218--226.

\bibitem{draxl2018nomad}
C.~Draxl, M.~Scheffler, Nomad: The fair concept for big data-driven materials science, Mrs Bulletin 43~(9) (2018) 676--682.

\bibitem{pizzi2016aiida}
G.~Pizzi, A.~Cepellotti, R.~Sabatini, N.~Marzari, B.~Kozinsky, Aiida: automated interactive infrastructure and database for computational science, Computational Materials Science 111 (2016) 218--230.

\bibitem{chanussot2021open}
L.~Chanussot, A.~Das, S.~Goyal, T.~Lavril, M.~Shuaibi, M.~Riviere, K.~Tran, J.~Heras-Domingo, C.~Ho, W.~Hu, et~al., Open catalyst 2020 (oc20) dataset and community challenges, Acs Catalysis 11~(10) (2021) 6059--6072.

\bibitem{choudhary2024jarvis}
K.~Choudhary, D.~Wines, K.~Li, K.~F. Garrity, V.~Gupta, A.~H. Romero, J.~T. Krogel, K.~Saritas, A.~Fuhr, P.~Ganesh, et~al., Jarvis-leaderboard: a large scale benchmark of materials design methods, npj Computational Materials 10~(1) (2024) 93.

\bibitem{zuo2020performance}
Y.~Zuo, C.~Chen, X.~Li, Z.~Deng, Y.~Chen, J.~Behler, G.~Cs{\'a}nyi, A.~V. Shapeev, A.~P. Thompson, M.~A. Wood, et~al., Performance and cost assessment of machine learning interatomic potentials, The Journal of Physical Chemistry A 124~(4) (2020) 731--745.

\bibitem{schmidt2024improving}
J.~Schmidt, T.~F. Cerqueira, A.~H. Romero, A.~Loew, F.~J{\"a}ger, H.-C. Wang, S.~Botti, M.~A. Marques, Improving machine-learning models in materials science through large datasets, Materials Today Physics 48 (2024) 101560.

\bibitem{kresse1996efficient}
G.~Kresse, J.~Furthm{\"u}ller, Efficient iterative schemes for ab initio total-energy calculations using a plane-wave basis set, Physical review B 54~(16) (1996) 11169.

\bibitem{kresse1996efficiency}
G.~Kresse, J.~Furthm{\"u}ller, Efficiency of ab-initio total energy calculations for metals and semiconductors using a plane-wave basis set, Computational materials science 6~(1) (1996) 15--50.

\bibitem{giannozzi2009quantum}
P.~Giannozzi, S.~Baroni, N.~Bonini, M.~Calandra, R.~Car, C.~Cavazzoni, D.~Ceresoli, G.~L. Chiarotti, M.~Cococcioni, I.~Dabo, et~al., Quantum espresso: a modular and open-source software project for quantumsimulations of materials, Journal of physics: Condensed matter 21~(39) (2009) 395502.

\bibitem{choudhary2021atomistic}
K.~Choudhary, B.~DeCost, Atomistic line graph neural network for improved materials property predictions, npj Computational Materials 7~(1) (2021) 185.

\bibitem{choudhary2023unified}
K.~Choudhary, B.~DeCost, L.~Major, K.~Butler, J.~Thiyagalingam, F.~Tavazza, Unified graph neural network force-field for the periodic table: solid state applications, Digital Discovery 2~(2) (2023) 346--355.

\bibitem{choudhary2024slmat}
K.~Choudhary, Slmat: A comprehensive serverless toolkit for advanced materials design (2024).

\bibitem{choudhary2024atomgpt}
K.~Choudhary, Atomgpt: Atomistic generative pretrained transformer for forward and inverse materials design, The Journal of Physical Chemistry Letters 15~(27) (2024) 6909--6917.

\bibitem{choudhary2018elastic}
K.~Choudhary, G.~Cheon, E.~Reed, F.~Tavazza, Elastic properties of bulk and low-dimensional materials using van der waals density functional, Physical Review B 98~(1) (2018) 014107.

\bibitem{choudhary2018computational}
K.~Choudhary, Q.~Zhang, A.~C. Reid, S.~Chowdhury, N.~Van~Nguyen, Z.~Trautt, M.~W. Newrock, F.~Y. Congo, F.~Tavazza, Computational screening of high-performance optoelectronic materials using optb88vdw and tb-mbj formalisms, Scientific data 5~(1) (2018) 1--12.

\bibitem{choudhary2019accelerated}
K.~Choudhary, M.~Bercx, J.~Jiang, R.~Pachter, D.~Lamoen, F.~Tavazza, Accelerated discovery of efficient solar cell materials using quantum and machine-learning methods, Chemistry of materials 31~(15) (2019) 5900--5908.

\bibitem{garrity2023fast}
K.~F. Garrity, K.~Choudhary, Fast and accurate prediction of material properties with three-body tight-binding model for the periodic table, Physical review materials 7~(4) (2023) 044603.

\bibitem{mandal2019systematic}
S.~Mandal, K.~Haule, K.~M. Rabe, D.~Vanderbilt, Systematic beyond-dft study of binary transition metal oxides, npj Computational Materials 5~(1) (2019) 115.

\bibitem{choudhary2021quantum}
K.~Choudhary, Quantum computation for predicting electron and phonon properties of solids, Journal of Physics: Condensed Matter 33~(38) (2021) 385501.

\bibitem{wines2023systematic}
D.~Wines, K.~Choudhary, F.~Tavazza, Systematic dft+ u and quantum monte carlo benchmark of magnetic two-dimensional (2d) crx3 (x= i, br, cl, f), The Journal of Physical Chemistry C 127~(2) (2023) 1176--1188.

\bibitem{wines2024combined}
D.~Wines, A.~Ibrahim, N.~Gudibandla, T.~Adel, F.~M. Abel, S.~Jois, K.~Saritas, J.~T. Krogel, L.~Yin, T.~Berlijn, et~al., A combined quantum monte carlo and dft study of the strain response and magnetic properties of two-dimensional (2d) 1t-vse $ \_2 $ with charge density wave, arXiv preprint arXiv:2409.19082 (2024).

\bibitem{pollard2025benchqc}
N.~Pollard, K.~Choudhary, Benchqc: A benchmarking toolkit for quantum computation, arXiv preprint arXiv:2502.09595 (2025).

\bibitem{choudhary2018machine}
K.~Choudhary, B.~DeCost, F.~Tavazza, Machine learning with force-field-inspired descriptors for materials: Fast screening and mapping energy landscape, Physical review materials 2~(8) (2018) 083801.

\bibitem{ward2018matminer}
L.~Ward, A.~Dunn, A.~Faghaninia, N.~E. Zimmermann, S.~Bajaj, Q.~Wang, J.~Montoya, J.~Chen, K.~Bystrom, M.~Dylla, et~al., Matminer: An open source toolkit for materials data mining, Computational Materials Science 152 (2018) 60--69.

\bibitem{tavazza2021uncertainty}
F.~Tavazza, B.~DeCost, K.~Choudhary, Uncertainty prediction for machine learning models of material properties, ACS omega 6~(48) (2021) 32431--32440.

\bibitem{choudhary2023atomvision}
K.~Choudhary, R.~Gurunathan, B.~DeCost, A.~Biacchi, Atomvision: A machine vision library for atomistic images, Journal of Chemical Information and Modeling 63~(6) (2023) 1708--1722.

\bibitem{choudhary2023chemnlp}
K.~Choudhary, M.~L. Kelley, Chemnlp: a natural language-processing-based library for materials chemistry text data, The Journal of Physical Chemistry C 127~(35) (2023) 17545--17555.

\bibitem{choudhary2024diffractgpt}
K.~Choudhary, Diffractgpt: Atomic structure determination from x-ray diffraction patterns using a generative pretrained transformer, The Journal of Physical Chemistry Letters 16 (2024) 2110--2119.

\bibitem{choudhary2025microscopygpt}
K.~Choudhary, Microscopygpt: Generating atomic-structure captions from microscopy images of 2d materials with vision-language transformers (2025).

\bibitem{choudhary2018high}
K.~Choudhary, A.~J. Biacchi, S.~Ghosh, L.~Hale, A.~R.~H. Walker, F.~Tavazza, High-throughput assessment of vacancy formation and surface energies of materials using classical force-fields, Journal of Physics: Condensed Matter 30~(39) (2018) 395901.

\bibitem{choudhary2017evaluation}
K.~Choudhary, F.~Y.~P. Congo, T.~Liang, C.~Becker, R.~G. Hennig, F.~Tavazza, Evaluation and comparison of classical interatomic potentials through a user-friendly interactive web-interface, Scientific data 4~(1) (2017) 1--12.

\bibitem{gale2003general}
J.~D. Gale, A.~L. Rohl, The general utility lattice program (gulp), Molecular Simulation 29~(5) (2003) 291--341.

\bibitem{salomon2013overview}
R.~Salomon-Ferrer, D.~A. Case, R.~C. Walker, An overview of the amber biomolecular simulation package, Wiley Interdisciplinary Reviews: Computational Molecular Science 3~(2) (2013) 198--210.

\bibitem{becker2013considerations}
C.~A. Becker, F.~Tavazza, Z.~T. Trautt, R.~A.~B. de~Macedo, Considerations for choosing and using force fields and interatomic potentials in materials science and engineering, Current Opinion in Solid State and Materials Science 17~(6) (2013) 277--283.

\bibitem{waters2023automated}
B.~Waters, D.~S. Karls, I.~Nikiforov, R.~S. Elliott, E.~B. Tadmor, B.~Runnels, Automated determination of grain boundary energy and potential-dependence using the openkim framework, Computational Materials Science 220 (2023) 112057.

\bibitem{jacobs2025practical}
R.~Jacobs, D.~Morgan, S.~Attarian, J.~Meng, C.~Shen, Z.~Wu, C.~Xie, J.~H. Yang, N.~Artrith, B.~Blaiszik, et~al., A practical guide to machine learning interatomic potentials--status and future, Current opinion in solid state materials science (2025).

\bibitem{wines2023high}
D.~Wines, K.~Choudhary, A.~J. Biacchi, K.~F. Garrity, F.~Tavazza, High-throughput dft-based discovery of next generation two-dimensional (2d) superconductors, Nano letters 23~(3) (2023) 969--978.

\bibitem{choudhary2021high}
K.~Choudhary, K.~F. Garrity, N.~J. Ghimire, N.~Anand, F.~Tavazza, High-throughput search for magnetic topological materials using spin-orbit spillage, machine learning, and experiments, Physical Review B 103~(15) (2021) 155131.

\bibitem{choudhary2022graph}
K.~Choudhary, T.~Yildirim, D.~W. Siderius, A.~G. Kusne, A.~McDannald, D.~L. Ortiz-Montalvo, Graph neural network predictions of metal organic framework co2 adsorption properties, Computational Materials Science 210 (2022) 111388.

\bibitem{choudhary2019high}
K.~Choudhary, K.~F. Garrity, F.~Tavazza, High-throughput discovery of topologically non-trivial materials using spin-orbit spillage, Scientific reports 9~(1) (2019) 8534.

\bibitem{choudhary2020computational}
K.~Choudhary, K.~F. Garrity, J.~Jiang, R.~Pachter, F.~Tavazza, Computational search for magnetic and non-magnetic 2d topological materials using unified spin--orbit spillage screening, NPJ Computational Materials 6~(1) (2020) 49.

\bibitem{choudhary2022designing}
K.~Choudhary, K.~Garrity, Designing high-tc superconductors with bcs-inspired screening, density functional theory, and deep-learning, npj Computational Materials 8~(1) (2022) 244.

\bibitem{wines2024data}
D.~Wines, K.~Choudhary, Data-driven design of high pressure hydride superconductors using dft and deep learning, Materials futures 3~(2) (2024) 025602.

\bibitem{wines2023inverse}
D.~Wines, T.~Xie, K.~Choudhary, Inverse design of next-generation superconductors using data-driven deep generative models, The Journal of Physical Chemistry Letters 14~(29) (2023) 6630--6638.

\bibitem{choudhary2017high}
K.~Choudhary, I.~Kalish, R.~Beams, F.~Tavazza, High-throughput identification and characterization of two-dimensional materials using density functional theory, Scientific reports 7~(1) (2017) 5179.

\bibitem{choudhary2020high}
K.~Choudhary, K.~F. Garrity, V.~Sharma, A.~J. Biacchi, A.~R. Hight~Walker, F.~Tavazza, High-throughput density functional perturbation theory and machine learning predictions of infrared, piezoelectric, and dielectric responses, npj computational materials 6~(1) (2020) 64.

\bibitem{choudhary2020data}
K.~Choudhary, K.~F. Garrity, F.~Tavazza, Data-driven discovery of 3d and 2d thermoelectric materials, Journal of Physics: Condensed Matter 32~(47) (2020) 475501.

\bibitem{johnson1999nist}
R.~D. Johnson~III, Nist 101. computational chemistry comparison and benchmark database (1999).

\bibitem{liu2015pdb}
Z.~Liu, Y.~Li, L.~Han, J.~Li, J.~Liu, Z.~Zhao, W.~Nie, Y.~Liu, R.~Wang, Pdb-wide collection of binding data: current status of the pdbbind database, Bioinformatics 31~(3) (2015) 405--412.

\bibitem{wines2024chips}
D.~Wines, K.~Choudhary, Chips-ff: Evaluating universal machine learning force fields for material properties, arXiv preprint arXiv:2412.10516 (2024).

\bibitem{wang2024examining}
S.-H. Wang, H.~Xin, L.~Achenie, K.~Choudhary, Examining generalizability of ai models for catalysis (2024).

\bibitem{choudhary2024intermat}
K.~Choudhary, K.~F. Garrity, Intermat: accelerating band offset prediction in semiconductor interfaces with dft and deep learning, Digital Discovery 3~(7) (2024) 1365--1377.

\bibitem{choudhary2023can}
K.~Choudhary, B.~G. Sumpter, Can a deep-learning model make fast predictions of vacancy formation in diverse materials?, AIP Advances 13~(9) (2023).

\bibitem{tran2016surface}
R.~Tran, Z.~Xu, B.~Radhakrishnan, D.~Winston, W.~Sun, K.~A. Persson, S.~P. Ong, Surface energies of elemental crystals, Scientific data 3~(1) (2016) 1--13.

\bibitem{zheng2020grain}
H.~Zheng, X.-G. Li, R.~Tran, C.~Chen, M.~Horton, D.~Winston, K.~A. Persson, S.~P. Ong, Grain boundary properties of elemental metals, Acta Materialia 186 (2020) 40--49.

\bibitem{olmsted2009survey}
D.~L. Olmsted, S.~M. Foiles, E.~A. Holm, Survey of computed grain boundary properties in face-centered cubic metals: I. grain boundary energy, Acta Materialia 57~(13) (2009) 3694--3703.

\end{thebibliography}





\end{document}